\documentclass[aps,twocolumn,showpacs,superscriptaddress]{revtex4}

\usepackage[dvips]{graphicx} \usepackage{amsmath} \usepackage{amsfonts}
\usepackage{amssymb}

\newcommand{\Tr}{\mathop{\mathrm{Tr}}} 
\newcommand{\iu}{i} 
\newcommand{\upd}{d} 
\newcommand{\omat}{\boldsymbol} 

\begin{document}

\title{Role of electronic structure in photoassisted transport
through atomic-sized contacts}

\author{J. K. Viljas}
\affiliation{Institut f\"ur Theoretische Festk\"orperphysik, 
Universit\"at Karlsruhe, D-76128 Karlsruhe, Germany}
\affiliation{Forschungszentrum Karlsruhe, 
Institut f\"ur Nanotechnologie, D-76021 Karlsruhe, Germany }

\author{J. C. Cuevas}
\affiliation{Departamento de F\'{\i}sica Te\'orica de la Materia
Condensada, Universidad Aut\'onoma de Madrid, E-28049 Madrid, Spain}
\affiliation{Institut f\"ur Theoretische Festk\"orperphysik, 
Universit\"at Karlsruhe, D-76128 Karlsruhe, Germany}
\affiliation{Forschungszentrum Karlsruhe, 
Institut f\"ur Nanotechnologie, D-76021 Karlsruhe, Germany }

\date{\today}

\begin{abstract}
  We study theoretically quantum transport through laser-irradiated
  metallic atomic-sized contacts. The radiation field is treated
  classically, assuming its effect to be the generation of an ac
  voltage over the contact. We derive an expression for the dc current
  and compute the linear conductance in one-atom thick contacts as a
  function of the ac frequency, concentrating on the role played by
  electronic structure. In particular, we present results for three
  materials (Al, Pt, and Au) with very different electronic
  structures. It is shown that, depending on the frequency and the
  metal, the radiation can either enhance or diminish the conductance.
  This can be intuitively understood in terms of the energy dependence
  of the transmission of the contacts in the absence of radiation.
\end{abstract}


\maketitle

\section{Introduction}
The study of electronic transport in microscopic and nanoscale
electrical contacts under the influence of time-dependent external
fields has a long history. Perhaps as the most famous example,
superconducting tunnel junctions subjected to microwave radiation
exhibit a step-like structure in their current-voltage ($I$-$V$)
characteristics \cite{Tucker85}. This can be understood in terms of
inelastic (``photoassisted'') transport of electrons across the
junction. In the early theoretical work of Tien and Gordon (TG)
\cite{Tien63}, this phenomenon was described by a harmonic voltage at
the radiation frequency $\omega$ applied to one of the leads, giving
rise to photo-``sidebands'' associated with the ``absorption'' or
``emission'' of an integer multiple of the photon energy
$\hbar\omega$. This theory has also been extended to describe
superconducting atomic point contacts \cite{Cuevas02} and the
predictions have been confirmed experimentally using microfabricated
Al break-junctions under microwave irradiation \cite{Chauvin06}. In
addition, a similar approach has been used to describe
laser-irradiated junctions in scanning tunneling microscopes (STM)
\cite{Grafstrom02}. In these, as a result of inherent asymmetries in
the geometry and the materials of the junction, laser irradiation can
cause dc (rectification) currents even in the absence of a dc bias
voltage \cite{Voelcker91,Levy91}.

Over the years, several types of model calculations have been employed
also to describe the ac-response of semiconductor heterostructures
\cite{Jauho94,Platero04} and other mesoscopic systems
\cite{Buttiker96,Pedersen98,Wang99}, as well as atomic and molecular
contacts \cite{Zheng00,Buker02,Kohler04,Galperin05,Wu05}.  Among
these, the TG-like theories have been quite successful in gaining a
qualitative understanding of light-induced currents
\cite{Tucker85,Platero04,Kohler04}.

\begin{figure}[!b]
\includegraphics[width=0.95\linewidth,clip=]{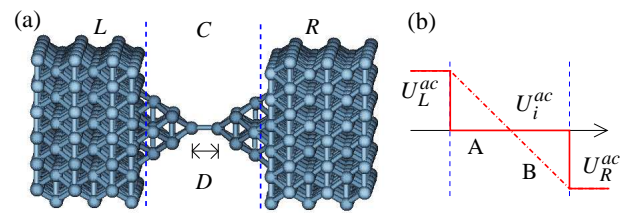}
\caption{(Color online) (a) Two infinite fcc [001] surfaces contacted 
with pyramids of 10 atoms in each, forming a ``dimer'' contact. The 
tip-to-tip distance is denoted with $D$, all other interatomic 
distances are the same as in bulk. (b) Two model ac voltage profiles 
$U_i^{ac}$, where $i$ labels atomic sites and orbitals in the $C$ 
region: a double step (A) and a linear ramp (B) between the lead 
values $U_{L}^{ac}$ and $U_{R}^{ac}$.}
\label{f.lcr}
\end{figure}

Metallic atomic-sized wires fabricated with STM or break-junction
techniques have turned out to be ideal systems for investigating
electronic transport at the nanoscale \cite{Agrait03}. The bulk of the
research in this field so far has concentrated on stationary transport
properties, while systems being driven by time-varying external fields
(such as laser light) have received less attention.  Very recent
experiments on laser-irradiated gold contacts support the idea that
photoassisted processes may play an important role in their transport
properties \cite{Guhr06}. To take the first steps towards a
microscopic description of experiments of this type, we address in
this paper the role of electronic structure in photoassisted
transport through atomic-sized junctions.  This problem is not only
relevant for the field of atomic contacts, but also for molecular
electronics, where the role of the metallic contacts on photoassisted
transport through molecular junctions remains to be understood.

In different types of metals, the nature and number of conduction
channels in one-atom contacts reflect the valence of the metal,
\emph{i.e.}, what type of atomic orbitals are available at the Fermi
energy \cite{Scheer98}. How is this difference between the metals seen
in their response to irradiation? To investigate this question, we use
a tight-binding model and explore mainly one-atom thick contact
geometries like the one shown in Fig.\ \ref{f.lcr}(a).  We are
interested in the linear conductance $G_{dc}=\partial I/\partial
V|_{V=0}$ as a function of the radiation frequency $\omega$, which we
call \emph{photoconductance}. Following the TG ideology, we model the
effect of radiation with a time-periodic voltage, but we also take
into account the effect of the voltage profile across the contact [see
Fig.\ \ref{f.lcr}(b)]. While many previous model calculations with
periodic driving fields are based on Floquet theory
\cite{Platero04,Kohler04}, our method is based on non-equilibrium
Green functions \cite{Levy91,Jauho94,Brandes97,Datta92}. Our approach
allows for a realistic description of the photoconductance of atomic
contacts, with a single free parameter describing the local intensity
of the radiation.

As examples of metals with very different electronic structures we
choose Al, Pt, and Au. We study one-atom thick junctions of these
materials mainly in the so-called contact regime, but also in the
tunneling regime. In the case of the contact regime we find the
following results. For Pt the effect of the ac voltage is almost
always to decrease $G_{dc}$ from its value in the absence of
radiation. This is because the Fermi energy $\epsilon_F$ lies at the
edge of the $d$ band, and exciting electrons above $\epsilon_F$
necessarily decreases their transmission probability, since less
transmission channels are available there. For Al, where $\epsilon_F$
is at the beginning of the $p$ band, also an enhancement of the
conductance is possible. The magnitude of these changes depends on the
intensity of the radiation, but can be up to $10\%$ or more at visible
frequencies. For Au, $\epsilon_F$ is in the $s$ band, with a single
completely open transmission channel for a wide range of energies.
Thus, low-frequency radiation has no effect on the conductance, while
in the visible range both an increase and a decrease are possible.
These conclusions are based on detailed numerical simulations, but
they can be understood in an appealing way in terms of the energy
dependence of the transmission in the absence of the ac drive.

The rest of the paper is organized as follows. In Sec.\ \ref{s.model}
we introduce the theoretical model used to describe the electronic
structure of atomic-sized contacts and the effect of irradiation.
Section \ref{s.formula} is devoted to the derivation of a general
formula for the dc current through atomic contacts subjected to an ac
voltage. In Sec.\ \ref{s.results} we describe examples of numerical
results of the frequency dependence of the linear conductance for
one-atom thick contacts of Al, Pt, and Au. Finally, in Sec.\
\ref{s.discussion} we discuss some of the assumptions and restrictions
of our model and present the main conclusions.

\section{Model}\label{s.model}
We describe the laser-irradiated point contact with an
$spd$-tight-binding Hamiltonian of the form
\begin{equation}
\begin{split} \label{e.hamilton}
H(t) & = H_{0} + H_{1}(t) \\
\hspace*{-0.5cm} H_{0} =\sum_{ij}d_i^\dagger H_{ij}d_j , &
\hspace*{0.5cm} H_{1}(t) = \sum_{ij}d_i^\dagger W_{ij}(t)d_j,
\end{split}
\end{equation}
where the indices $i,j$ run over the different atoms and orbitals,
including the two spin directions. The creation and annihilation
operators ($d^\dagger_i$ and $d_j$) satisfy $\{d_i,d_j\}=0$ and
$\{d_i,d^\dagger_j\}=[\omat{S}^{-1}]_{ij}$, where $\omat{S}$ is the
overlap matrix of the non-orthogonal basis \cite{Viljas05}. The
Hamiltonian $H_{0}$ is for the system without radiation and dc
voltage, while $H_{1}(t)$ includes them both.  The matrix elements of
the Hamiltonian and the overlap matrix ($H_{ij}$ and $S_{ij}$) are
taken from the parametrization of Ref.\ \onlinecite{Papa}.  All
matrices in the spin-orbital basis (such as $\omat{H}$ and $\omat{S}$)
are denoted with a boldface symbol.

We consider ideal symmetric geometries of the type shown in Fig.\
\ref{f.lcr}(a), with a single-atom thick constriction.  This type of
``dimer'' structure is suggested by molecular dynamics simulations as
the most common one in the last conductance plateau \cite{Dreher05}.
When the distance $D$ between the tip atoms corresponds to the bulk
interatomic distance, the junction is said to be in the contact
regime. We also study larger $D$ values, where the junction enters the
tunneling regime. For the calculation of transport we shall, in the
usual way, divide the system into left lead ($L$), central ($C$), and
right lead ($R$) regions. The leads are modeled with infinite
surfaces, where the fcc [001] axis coincides with the transport
direction. We index the regions with the label $\Omega=L,C,R$.  As a
matrix index, this label also indicates collectively all the orbital
indices of the respective region \cite{Viljas05}.

We assume that the $L$ and $R$ lead potentials are spatially constant
and harmonic with angular frequency $\omega$, such that
$T_\omega=2\pi/\omega$ is the oscillation period. Thus
$\omat{W}_{\Omega\Omega}(t)=U_\Omega(t)\omat{S}_{\Omega\Omega}$, where
$U_{\Omega}(t)=U^{dc}_{\Omega}+U^{ac}_{\Omega}\cos(\omega t)$, and
$\Omega=L,R$. The applied dc voltage $V=(U_L^{dc}-U_R^{dc})/e$, where
$-e$ is the electron charge, is assumed to be infinitesimal.  For the
lead-center hoppings we also assume
$\omat{W}_{C\Omega}(t)=U_{\Omega}(t)\omat{S}_{C\Omega}$, etc.  The
central potential is assumed to be of the form
$\omat{W}_{CC}(t)=\omat{W}_{CC}^{dc}+\omat{W}_{CC}^{ac}\cos(\omega
t)$, or $[\omat{W}_{CC}(t)]_{ij} = [\omat{S}_{CC}]_{ij}
(U_i(t)+U_j(t))/2$ \cite{Note1}.  Here,
$U_{i}(t)=U_i^{dc}+U_i^{ac}\cos(\omega t)$ is the same for all
orbitals $i$ on the same atom in region $C$.  The actual shape of
$U_{i}(t)$ within $C$ should in principle be obtained
self-consistently through the solution of a Poisson equation, such
that screening and local field-enhancement effects would be properly
accounted for.  In this case the appearance of higher harmonics of
$\omega$ and phase shifts in $\omat{W}_{CC}(t)$ would be possible.  As
this would be computationally very costly, we shall do the following:
the dc part $U_i^{dc}$ is fixed by a requirement of local charge
neutrality in equilibrium \cite{Viljas05,Note1} and incorporated into
$H_{CC}$, while for $U_i^{ac}$ we just assume some simple forms.
Below we consider two model profiles [see Fig.\ \ref{f.lcr}(b)]:
$U_i^{ac}=(U_L^{ac}+U_R^{ac})/2$ for all orbitals $i$ in region $C$
(profile A), and one which linearly interpolates between $U_L^{ac}$
and $U_R^{ac}$ (profile B).  For the symmetric junctions that we are
considering, symmetric profiles of this type are the most reasonable.

\section{Current formula}\label{s.formula}
We only consider the time-averaged current, and can thus neglect
displacement contributions without losing current conservation
\cite{Pedersen98,Buttiker96,Wang99}. The particle current is given in
terms of non-equilibrium Green functions by \cite{Jauho94}
\begin{equation}
\begin{split}\label{e.curr1}
I &= \frac{e}{\hbar}\int_0^{T_\omega}\frac{\upd t}{T_\omega}
\Tr[(\omat{G}^{<}_{CL}\circ \omat{t}_{LC}
-\omat{t}_{CL}\circ \omat{G}^{<}_{LC})(t,t)] . \\
\end{split}
\end{equation}
Here, in the case of the non-orthogonal basis \cite{Viljas05}
\begin{equation}
\begin{split} \label{e.hoppings}
\omat{t}_{CL}(t,t')&=[(\omat{H}_{CL}+\omat{W}_{CL}(t))
-\omat{S}_{CL}\iu\hbar\partial_t]\hbar\delta(t-t') , \nonumber
\end{split}
\end{equation}
and using the ``Langreth rules'', $\omat{G}^{<}_{CL} =
\omat{G}^{r}_{CC}\circ \omat{t}_{CL}\circ \omat{g}^{<}_{LL}
+\omat{G}^{<}_{CC}\circ \omat{t}_{CL}\circ \omat{g}^{a}_{LL}$
\cite{Jauho94}.  The product $\circ$ is defined by $(\omat{A}\circ
\omat{B})(t,t')=\int(\upd s/\hbar)\omat{A}(t,s)\omat{B}(s,t')$, where
$\omat{A}$ and $\omat{B}$ are matrices in the spin-orbital basis, over
which the trace $\Tr$ acts.  The Green functions
$\omat{G}^x_{\Omega\Omega'}$ with $x=r,a,\gtrless$ and
$\Omega,\Omega'=L,C,R$ are defined as usual \cite{Jauho94,Viljas05},
and $\omat{g}^{x}_{\Omega\Omega}$ are the functions for uncoupled
leads.  The $CC$ component of the full retarded function satisfies
\begin{equation}
\begin{split} \label{e.gcc}
[\omat{S}_{CC}\iu\hbar \frac{\partial}{\partial t} -\omat{H}_{CC}-\omat{W}_{CC}(t)]
\omat{G}^{r}_{CC}(t,t') =\hbar \delta(t-t') \\
+(\omat{\Sigma}^r_{L}\circ \omat{G}^{r}_{CC})(t,t') +
(\omat{\Sigma}^r_{R}\circ \omat{G}^{r}_{CC})(t,t') ,
\end{split}
\end{equation}
whereas
$\omat{G}^{\gtrless}_{CC}=\omat{G}^r_{CC}\circ(\omat{\Sigma}^{\gtrless}_L
+\omat{\Sigma}^{\gtrless}_R)\circ \omat{G}^a_{CC}$. Here we defined the 
``lead self-energies'' $\omat{\Sigma}^x_{\Omega} 
= \omat{t}_{C\Omega}\circ \omat{g}^x_{\Omega\Omega}\circ \omat{t}_{\Omega C}$, 
with $\Omega=L,R$. The solutions of 
\begin{equation}
\begin{split}
[\omat{S}_{\Omega\Omega}\iu\hbar\frac{\partial}{\partial t} 
-\omat{H}_{\Omega\Omega}-\omat{W}_{\Omega\Omega}(t)]
\omat{g}^r_{\Omega\Omega}(t,t') & = \hbar\delta(t-t') 
\end{split}
\end{equation}
lead to self-energies of the form
\begin{equation}\begin{split}
\omat{\Sigma}^x_\Omega(t,t') = 
e^{-\iu\int_{t'}^t\frac{\upd s}{\hbar} U_\Omega(s)}
\int\frac{\upd\epsilon}{2\pi} e^{-\iu\epsilon(t-t')/\hbar}
\omat{\Sigma}^{x,eq}_\Omega(\epsilon) .
\end{split}\end{equation}
The equilibrium self-energies are given by
$\omat{\Sigma}^{x,eq}_\Omega(\epsilon)=\omat{t}_{C\Omega}(\epsilon)
\omat{g}^{x,eq}_{\Omega\Omega}(\epsilon)\omat{t}_{\Omega C}(\epsilon)$, 
where the Green functions
$\omat{g}^{x,eq}_{\Omega\Omega}(\epsilon)$ of the infinite surfaces
are obtained by a decimation method \cite{Viljas05}, while
$\omat{t}_{C\Omega}(\epsilon)=\omat{H}_{C\Omega}-\epsilon
\omat{S}_{C\Omega}$ and $\omat{t}_{\Omega C}=[\omat{t}_{C\Omega}]^\dagger$. 
We also have
$\omat{g}^{<,eq}_{\Omega\Omega}(\epsilon)=-f(\epsilon)$
$[\omat{g}^{r,eq}_{\Omega\Omega}(\epsilon)-\omat{g}^{a,eq}_{\Omega\Omega}(\epsilon)]$
where $f(\epsilon)$ is the Fermi function.

Since the Green functions and self-energies $\omat{A}(t,t')$ are all 
periodic in the time $(t+t')/2$, we can simplify the analysis by 
working in Fourier coordinates. We define the harmonic matrices 
$\hat{A}(\epsilon)$ with components
$[\hat A]_{m,n}(\epsilon)=\omat{A}_{m-n}(\epsilon+(m+n)\hbar\omega/2)$, 
where $m,n$ are integers, and
\begin{equation}\begin{split}
\omat{A}_n(\epsilon) &= \int_0^{T_\omega} 
\frac{\upd T}{T_\omega} e^{\iu n\omega T}
\int\frac{\upd \tau}{\hbar}e^{\iu\epsilon \tau/\hbar}
\omat{A}(T+\tau/2,T-\tau/2) . \nonumber
\end{split}\end{equation}
In this way we find
$\hat{G}_{}^{\gtrless}=\hat{G}_{}^r(\hat{\Sigma}^{\gtrless}_L 
+\hat{\Sigma}^{\gtrless}_R)\hat{G}_{}^a$,
while
\begin{equation}\begin{split}
[\hat{G}_{}^{r,a}]^{-1}
&=[\hat\epsilon S_{CC}-H_{CC} ]\hat{1}
-\hat W_{} -\hat{\Sigma}^{r,a}_L-\hat{\Sigma}^{r,a}_R . 
\end{split}\end{equation}
Here $[\hat\epsilon]_{m,n}=(\epsilon+m\hbar\omega)\delta_{m,n}$,
$[\hat W_{}]_{m,n}=\omat{W}_{CC}^{ac}(\delta_{m-1,n}+\delta_{m+1,n})/2$,
and we have dropped the indices $CC$ from the harmonic matrices.
The self-energies as well as the scattering rates $\hat{\Gamma}_{\Omega}=
\iu(\hat{\Sigma}^r_\Omega-\hat{\Sigma}^a_\Omega)$ 
are related to the corresponding equilibrium quantities by
\begin{equation}
\begin{split}
[\hat\Sigma^x_\Omega]_{m,n}
&=\sum_{l} [\hat\Sigma^{x(l)}_\Omega]_{m,n}, \quad
[\hat\Gamma_\Omega]_{m,n}
=\sum_{l} [\hat\Gamma^{(l)}_\Omega]_{m,n} , 
\end{split}
\end{equation}
where we define the components
\begin{equation}\begin{split}
[\hat\Gamma^{(l)}_\Omega]_{m,n}(\epsilon) &=
J_{m+l}\left(\alpha_\Omega\right)
J_{n+l}\left(\alpha_\Omega\right)
\omat{\Gamma}_\Omega^{eq}(\epsilon-U_\Omega^{dc}-l\hbar\omega) , \nonumber
\end{split}\end{equation}
with a similar equation for $\hat{\Sigma}^{x(l)}_\Omega(\epsilon)$.
Here 
$\omat{\Gamma}_\Omega^{eq}(\epsilon)
=\iu[\omat{\Sigma}^{r,eq}(\epsilon)-\omat{\Sigma}^{a,eq}(\epsilon)]$,
$J_l$ are Bessel functions of the first kind,
and $\alpha_\Omega=U_\Omega^{ac}/\hbar\omega$. 
All of the harmonic matrices satisfy the symmetry
$[\hat A^{(l)}]_{m+k,n+k}(\epsilon) = 
[\hat A^{(l+k)}]_{m,n}(\epsilon+k\hbar\omega)$.
Finally, we may note that
$\int(\upd t/T_\omega)\Tr[(\omat{A}\circ \omat{B})(t,t)]
=\int_0^{\hbar\omega}(\upd\epsilon/2\pi)\Tr_{\omega}
[\hat{A}(\epsilon)\hat{B}(\epsilon)]$,
where ${\Tr}_{\omega}[\hat{A}(\epsilon)]=\sum_m\Tr [\hat{A}]_{m,m}(\epsilon)$.
For numerical calculations, the matrices $\hat A(\epsilon)$ must be 
truncated to a few lowest indices $m,n$, but the results converge 
rapidly with the cutoff.

Using the above definitions and general symmetries like 
$\hat{G}^r-\hat{G}^a=\hat{G}^>-\hat{G}^<$, 
Eq.\ (\ref{e.curr1}) yields the dc current
\begin{equation}\begin{split}\label{e.elcurr2}
I &= \frac{e}{\hbar}
\int_0^{\hbar\omega}\frac{\upd\epsilon}{2\pi}
\sum_{k,l} 
{\Tr}_\omega[\hat{G}_{}^r\hat{\Gamma}_R^{(k)}
\hat{G}_{}^a\hat{\Gamma}_L^{(l)}]
(f_L^{(l)}-f_R^{(k)}) ,
\end{split}\end{equation}
where
$f_{\Omega}^{(k)}(\epsilon) = f(\epsilon-U_\Omega^{dc}-k\hbar\omega)$.
In the absence of an ac field this reduces to the standard Landauer-type 
formula \cite{Datta95}. Although we are not computing $U_i(t)$ 
self-consistently, Eq.\ (\ref{e.elcurr2}) is still gauge invariant in 
the sense that a spatially constant potential added everywhere has no 
effect. Thus, the results only depend on $U_L(t)-U_R(t)$ \cite{Pedersen98}.

\section{Results}\label{s.results}
The experimentally accessible quantity which we calculate
is the linear conductance $G_{dc}(\omega)=\partial I/\partial V|_{V=0}$. 
In what follows we shall assume zero temperature. If the
lead self-energies $\Sigma^{r,eq}_{L,R}$
are furthermore assumed to be energy-independent (``wide-band'' 
approximation), then the full result for potential profile A simplifies to 
\cite{Pedersen98}
\begin{equation}\begin{split} \label{e.simpleg}
G_{dc}(\omega) = 
G_0 \sum_l [J_l(\alpha/2)]^2 T^{eq}(\epsilon_F + l\hbar\omega) ,
\end{split}\end{equation}
where $G_0=2e^2/h$ and $T^{eq}(\epsilon)$ is the equilibrium transmission 
function \cite{Note2}.
We have also defined the parameter 
$\alpha=\alpha_L-\alpha_R=(U^{ac}_L-U^{ac}_R)/\hbar\omega$, 
which measures the local intensity of the radiation \cite{Tien63}.
Equation (\ref{e.simpleg}) describes electrons incident from the 
different sidebands (after having ``absorbed'' or ``emitted'' $l$ photons)
being transmitted elastically through the constriction, which mostly 
determines $T^{eq}(\epsilon)$.
In reality, the electric field is nonzero only in the constriction
and thus the actual physical transitions must occur there. 
Note that Eq.\ (\ref{e.simpleg}) should reproduce the 
results for profile A only in the limit of small $\omega$
and $\alpha$. Still, it works surprisingly well for all the 
cases presented below.

\begin{figure}[!tb]
\includegraphics[width=0.95\linewidth,clip=]{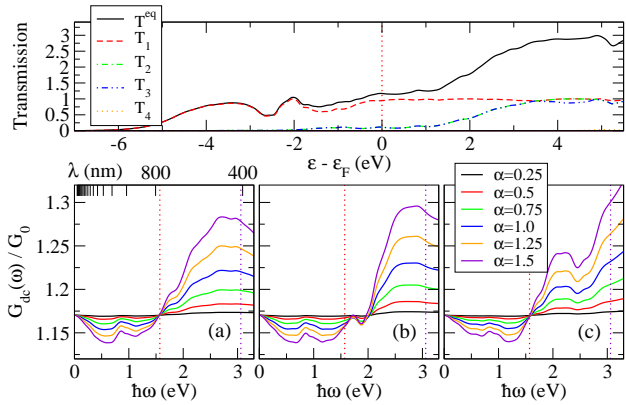}
\caption{(Color online) Top panel: Equilibrium transmission $T^{eq}$ 
and its decomposition into conduction channels $T_{1,2,3,4}$ for an Al dimer 
contact. The position of $\epsilon_F$ is indicated by a vertical dotted line.
Lower panels: Zero-temperature photoconductance for several values of
$\alpha$ as a function of frequency $\omega$ using the voltage 
profile A (a), profile B (b) and Eq.\ (\ref{e.simpleg}) (c). 
In (b) the wavelengths $\lambda$ with a tick spacing of 400 nm are shown.
The range of visible light is indicated by vertical dotted lines.
}
\label{f.aluminum}
\end{figure}
Let us first discuss the results for the contact regime, which 
is the subject of our main interest in this paper.
In Fig.\ \ref{f.aluminum} we show results for a dimer Al
contact, where $\epsilon_F$ lies in the $3p$ band. 
The upper panel shows the transmission function $T^{eq}(\epsilon)$
and the lower panels show $G_{dc}(\omega)$ for the two voltage profiles
as well Eq.\ (\ref{e.simpleg}) for several values of $\alpha$. 
In the absence of an ac voltage ($\omega=0$) the conductance 
is close to $G_0$, and is dominated by three conduction channels 
due to the contribution of $3s$ and $3p$ orbitals \cite{Scheer98}. 
At finite $\omega$, the relative change 
$\delta G_{dc}(\omega)=[G_{dc}(\omega)-G_{dc}(\omega=0)]/G_{dc}(\omega=0)$ 
is initially \emph{negative}, but can then rise to \emph{positive} values of 
$\approx 10\%$ towards visible frequencies. This behavior is similar 
for both profiles A and B, as well as for Eq.\ (\ref{e.simpleg}). From 
the latter result, we can interpret our findings in the following 
appealing way. For $\alpha<1$ only the first sidebands ($l=0,\pm 1$)
contribute to the transport. In this limit, according to Eq.\ 
(\ref{e.simpleg}), $\delta G_{dc}(\omega)$ measures, roughly speaking, 
the ``second derivative'' of $T^{eq}(\epsilon)$ on the scale of 
$\hbar\omega$ around $\epsilon_F$. Thus, for instance, the conductance 
enhancement in the visible range follows from the transmission increase 
for electrons promoted above $\epsilon_F$ (due to ``absorption'') 
overcoming the corresponding decrease for electrons moved below 
$\epsilon_F$ (due to ``emission'').

\begin{figure}[!t]
\includegraphics[width=0.95\linewidth,clip=]{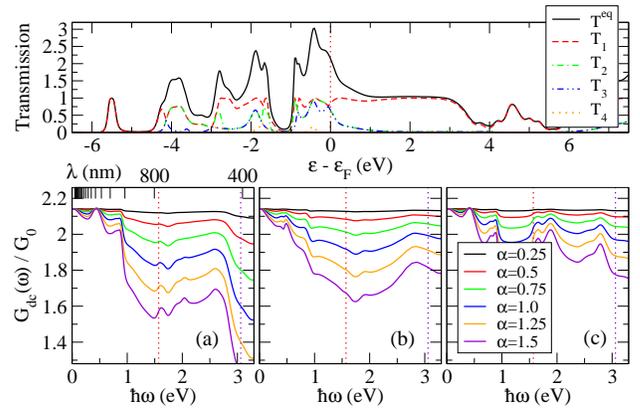}
\caption{ (Color online)
Same as Fig.\ \ref{f.aluminum}, but for Pt.
}
\label{f.platinum}
\end{figure}

Figure \ref{f.platinum} shows the corresponding results for Pt.
In the absence of radiation the conductance is close to $2.1G_0$
due to the contributions of mainly three conduction channels, which
originate from the $6s$ and $5d$ orbitals.
In this case, and in the contact regime in general for Pt, the 
effect of the radiation is almost always a significant reduction 
in conductance. This is understandable, since $\epsilon_F$ lies at 
the edge of the $d$ band, and photon absorption leads to an energy 
region where fewer open transmission channels are available and $T^{eq}$ 
is smaller. Note that for low $\omega$, the full results are again 
well described by Eq.\ (\ref{e.simpleg}). Let us remark that we only 
compute $G_{dc}(\omega)$ for low enough $\omega$ and $\alpha$, so that the 
electric fields of the radiation remain reasonable 
($\lesssim 3\cdot 10^9$ V/m).

\begin{figure}[!t]
\includegraphics[width=0.95\linewidth,clip=]{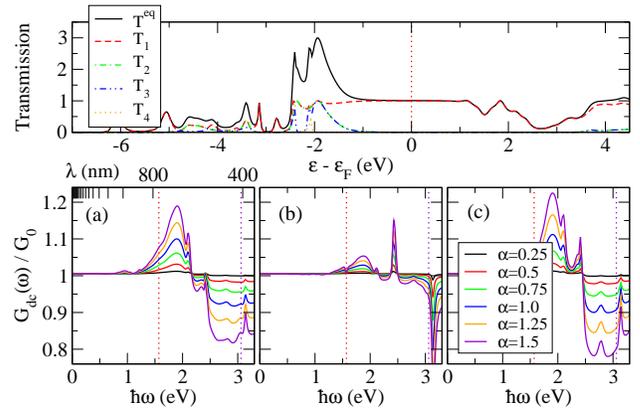}
\caption{ (Color online)
Same as Fig.\ \ref{f.aluminum}, but for Au.
}
\label{f.gold}
\end{figure}

The results for Au are shown in Fig.\ \ref{f.gold}. As can be seen
in the upper panel, the conductance for $\omega=0$ is equal to 
$1G_0$ with a single open channel arising from the contribution 
of the $6s$ orbitals \cite{Scheer98}. Moreover, notice that 
the transmission around $\epsilon_F$ is very flat. Due to this flatness, 
for frequencies up to $\hbar\omega\approx 1$ eV the effect of radiation 
is practically negligible. In the red part of the visible 
range ($\hbar\omega\lesssim 2$ eV) we find that $\delta G_{dc}(\omega)>0$ 
up to a few percent, although this depends rather strongly on the
choice of the voltage profile. This increase in the conductance 
is due to a contribution of the $5d$ bands located $2$ eV below 
$\epsilon_F$, where the number of open transmission channels is 
higher than at $\epsilon_F$. At higher frequencies 
$\delta G_{dc}(\omega)<0$, as for Pt. We have also studied 
Au contacts with atomic chains of varying length and the results 
remain qualitatively similar, although in the case of profile B 
the amplitude $\delta G_{dc}(\omega)<0$ becomes smaller as
the number of chain atoms increases. 

\begin{figure}[!t]
\includegraphics[width=0.95\linewidth,clip=]{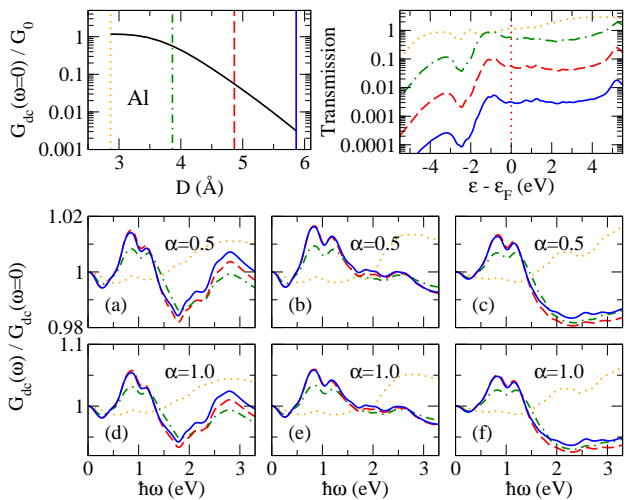}
\caption{ (Color online)
Tunneling limit for Al contacts. 
Top left panel: zero-frequency conductance $G_{dc}(\omega=0)$
as a function of tip distance $D$. The vertical lines indicate
the distances where the examples with corresponding linestyles 
in the other panels are computed. Top right panel: transmission 
$T^{eq}(\epsilon)$ for the example distances. Lower panels: 
the left-hand (a,d), central (b,e),
and right-hand (c,d) panels are for profile A, profile B, and 
Eq.\ (\ref{e.simpleg}), respectively. The panels (a)-(c) are for
$\alpha=0.5$ and (d)-(f) for $\alpha=1.0$.
}
\label{f.altunnel}
\end{figure}

When the distance $D$ between the tip atoms of the dimer contact
(Fig.\ \ref{f.lcr}) is increased, the tunneling regime is approached.
Here $G_{dc}(\omega)$ decreases exponentially with increasing $D$, but
$\delta G_{dc}(\omega)$ has a tendency to saturate. This is easy to
understand, because it may be shown that for very large $D$ the
magnitude of the conductance is approximately determined by the square
of the slowest-decaying hopping integral between the tip atoms, which
enters the conductance formula as a prefactor. The form of $\delta
G_{dc}(\omega)$ can, however, be very different from the contact
regime.

\begin{figure}[!t]
\includegraphics[width=0.95\linewidth,clip=]{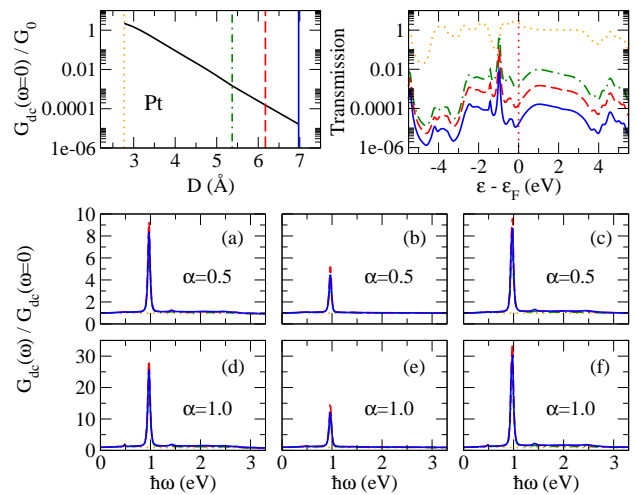}
\caption{ (Color online)
Same as Fig.\ \ref{f.altunnel}, but for Pt.}
\label{f.pttunnel}
\end{figure}

\begin{figure}[!t]
\includegraphics[width=0.95\linewidth,clip=]{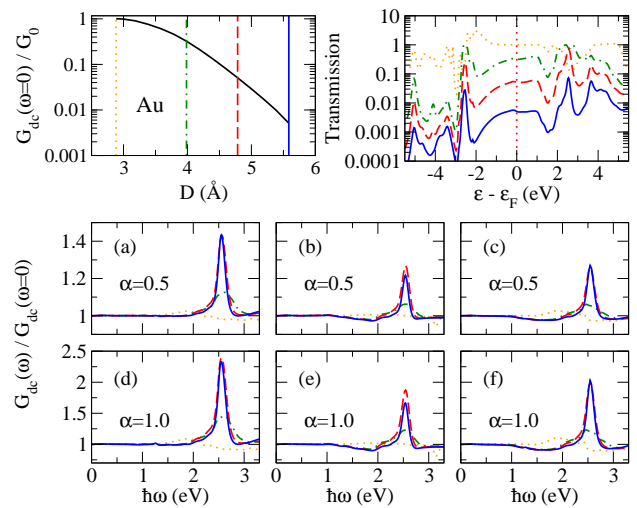}
\caption{ (Color online)
Same as Fig.\ \ref{f.altunnel}, but for Au.}
\label{f.autunnel}
\end{figure}

An example of the tunneling regime results for Al is presented in
Fig.\ \ref{f.altunnel}. The top panels illustrate the exponential
decay of $G_{dc}(\omega=0)$ and $T^{eq}(\epsilon)$ with $D$, while in
the lower panels the quantity $G_{dc}(\omega)/G_{dc}(\omega=0)$ is
shown for two values of $\alpha$ and for several distances $D$. The
results again look quite similar for the two voltage profiles as well
as for Eq.\ (\ref{e.simpleg}). The quantity $\delta G_{dc}$ can obtain
both positive and negative values, and its saturation with $D$ is
clearly visible. For Pt (see Fig.\ \ref{f.pttunnel}) we find that
$G_{dc}(\omega)$ is otherwise flat, but there is a sharp resonance at
$\hbar\omega\approx1$ eV where $\delta G_{dc}(\omega)$ can take on
\emph{positive} values of up to a few hundred percent. This is due to
resonant transmission through a level formed by the $d$ orbitals of
the tip atoms, which can be seen also in the $T^{eq}(\epsilon)$
curves.  In the case of Au there exists a rather similar, but broader
positive peak covering the visible range (see Fig.\
\ref{f.autunnel}). For each metal, we only consider large enough $D$
to see the saturation of $\delta G_{dc}(\omega)$.  Indeed, the
tight-binding parametrization we are using is based on bulk
calculations \cite{Papa} and is probably not good for very large
interatomic distances. Furthermore, the charge neutrality shifts
mentioned in Sec.\ \ref{s.model} are strongest for the tip atoms.
Therefore the peaks in $G_{dc}(\omega)$ observed for Pt and Au, for
example, should be taken with some reservations.  For very large $D$,
we would also not expect to see such a good agreement between the
results for profiles A and B, because the tunneling conductance for
profile B depends on the local densities of states of the tip atoms in
a way that cannot be written in the form (\ref{e.simpleg}).
Nevertheless, our results serve as good illustrations of the different
phenomena that may potentially arise in the tunneling regime.

Above we have only shown some example results for very idealized
symmetric geometries. In general, the signs and magnitudes of $\delta
G_{dc}(\omega)$ depend sensitively on details of the atomic structure,
as does $T^{eq}(\epsilon)$. For a more complete analysis one should
therefore study also larger contacts and carry out a statistical
exploration of geometrical variations along the lines of Ref.
\onlinecite{Dreher05}. Based on Eq. (\ref{e.simpleg}), we can still
expect that in the limit of several-atoms-wide contacts the relative
effect of the ac voltage gradually becomes smaller as
$T^{eq}(\epsilon)$ becomes smoother. A direct comparison with the
ongoing experiments \cite{Guhr06} will be postponed for later.

\section{Discussion and conclusions}\label{s.discussion}
Some of the assumptions of our model and the effects not taken into
account are worth discussing. First, we assume a flat potential in the
leads, which requires a complete screening of the electric field. This
is well satisfied in metals for frequencies $\omega$ much below the
plasma frequency but, as $\omega$ begins to approach the visible
range, the screening is weakened. On the other hand, as we have
mentioned, the screening in the central region is not treated
self-consistently.  One of the main concerns here is that local
surface-plasmon modes in small geometries tend to have their
frequencies close to the visible range, and their excitation can lead
to huge field-enhancement effects \cite{Grafstrom02}. Although this is
not a problem for our model, one should bear in mind that $\alpha$ is
not simply related to the field intensity away from the contact and,
following the spirit of the experiments in superconducting contacts
\cite{Chauvin06}, it must be understood as an adjustable parameter.
Also, at visible frequencies ``multiphoton'' processes can already
cause a photoelectric effect, \emph{i.e.}, excite electrons above the
vacuum level, which typically lies $4$--$6$ eV above $\epsilon_F$.

Heating accompanies any possible effect arising from absorption of
light, which in metals is specially pronounced in the optical range
due to the onset of interband transitions. While we may expect that
the effect of temperature is just to broaden our results, in practice
thermal expansion can play an important role.  This is well documented
in the STM context \cite{Grafstrom02}, where expansion typically
brings the tip closer to the sample, thus reducing the tunneling gap
width. This results in a strong enhancement of the tunneling current.
In the case of atomic wires (\emph{i.e.}, in the contact regime), it
is not obvious in which sense and to what degree thermal expansion
affects the conductance.  Assuming that the expansion simply mimics a
mechanical closing process in the STM or break-junction experiments,
it can lead to either an increase or a decrease of the conductance,
depending on the metal. For instance, for Al contacts, which exhibit
raising plateaus upon stretching \cite{Cuevas1998}, one would expect a
decrease of the conductance due to thermal expansion, as opposed to
the effect of the electronic structure in the visible range (see Fig.\
\ref{f.aluminum}). In this sense, our predictions can be valuable for
distinguishing in an experiment (such as Ref.\ \onlinecite{Guhr06})
between the contributions of the different effects to the
photoconductance.

In conclusion, we have modeled electronic transport in atomic point
contacts subjected to external electromagnetic radiation.  The
radiation has been described by an ac voltage over the contact.
Within a non-equilibrium Green function method, we have derived a
formula for the dc current in the presence of such an ac drive.  Using
a tight-binding model, we have applied the method for describing
atomic-sized contacts of Al, Pt, and Au, and have found that the
qualitative modification of the dc conductance by the ac voltage can
be predicted from the equilibrium transmission function. Depending on
the metal, the detailed structure of the contact, and the external
frequency, the effect can be either an increase or a decrease in the
conductance. At present, experiments are under way to test these
predictions \cite{Guhr06}.

\acknowledgments

We acknowledge useful discussions with D. Schmidt, E. Scheer, 
P. Leiderer, R. H. M. Smit, F. Pauly, S. Wohlthat, and M. H\"afner. 
This work was financially supported by the Helmholtz Gemeinschaft 
(Contract No.\ VH-NG-029) and by the DFG within the Center for Functional 
Nanostructures.

\end{document}